\documentclass[fleqn,twoside]{article}
\usepackage{espcrc2_mod}

\usepackage{graphicx}
\usepackage[figuresright]{rotating}

\newcommand{\AmS}{{\protect\the\textfont2
  A\kern-.1667em\lower.5ex\hbox{M}\kern-.125emS}}

\begin{document}

\thispagestyle{empty}
\title{\begin{flushright}SLAC-PUB-13831\\\end{flushright}Hamiltonian light-front field theory within an AdS/QCD basis}

\author{J.~P.~Vary\address[ISU]{Department of Physics and Astronomy, Iowa 
State University, Ames, Iowa 50011, USA}%
        \thanks{This work is supported in part by a DOE Grant DE-FG02-87ER40371 and by 
DOE Contract 
DE-AC02-76SF00515.},
H.~Honkanen\addressmark[ISU],
Jun~Li\addressmark[ISU],
P.~Maris\addressmark[ISU],
S.~J.~Brodsky\address{SLAC National Accelerator Laboratory, Stanford 
University, Menlo Park, California, USA},
A.~Harindranath\address{Theory Group, Saha Institute of Nuclear Physics,1/AF, 
Bidhannagar, Kolkata, 700064, India},
G.~F.~de~Teramond\address{Universidad de Costa Rica, San Jos\'e, Costa Rica},
P.~Sternberg\address[LBL]{Lawrence Berkeley National Laboratory, Berkeley, 
California, USA}
\thanks{Currently at ILOG Inc, Incline Village, NV}, 
E.~G.~Ng\addressmark[LBL],
C.~Yang\addressmark[LBL]}

\begin{abstract}

Non-perturbative Hamiltonian light-front quantum field theory presents 
opportunities and challenges that bridge particle physics and nuclear physics.
Fundamental theories, such as Quantum Chromodynmamics (QCD) and
Quantum Electrodynamics (QED) offer the promise of great predictive power 
spanning phenomena on all scales from the microscopic to cosmic scales, but 
new tools that do not rely exclusively on perturbation theory are required
to make connection from one scale to the next. We outline recent theoretical 
and computational progress to build these bridges and provide illustrative 
results for nuclear structure and quantum field theory.
As our framework we choose  light-front gauge and a basis function 
representation with two-dimensional harmonic oscillator basis for transverse
modes that corresponds with eigensolutions of the soft-wall AdS/QCD model 
obtained from light-front holography.

\vspace{1pc}
\end{abstract}

\maketitle

\section{Introduction}

Major goals of Hamiltonian light-front field theory include predicting both 
the masses and transitions rates of the
hadrons and their structures as seen in high-momentum transfer experiments.
So far, the approaches 
in a discretized momentum basis \cite{BPP98} and in transverse lattice
\cite{BD2002,JPV214,GP08} have shown significant promise.
The main advantages of Hamiltonian light-front quantum field theory are: (1)
one  evaluates experimental observables that are non-perturbative and 
relativistically invariant such as masses, form factors, 
structure functions, etc.; (2) one evaluates these quantities in 
Minkowski space; and (3) there is no fermion doubling problem.

Here we present  a basis-function approach that exploits
recent advances in solving the non-relativistic strongly interacting nuclear
many-body problem \cite{NCSM,NCFC}.  Both light-front field 
theory and nuclear many-body theory face common issues within the Hamiltonian 
approach - i.e. how to 
(1) define the Hamiltonian;
(2) renormalize to a finite space;
(3) solve for non-perturbative observables while preserving as many 
symmetries as possible; and,
(4) take the continuum limit.
Nevertheless, Ken Wilson has assessed the advantages of 
adopting advances in quantum many-body theory and has long advocated adoption 
of basis function methods as an alternative to the lattice gauge approach 
\cite{KW1989}.

\section{Ab initio Hamiltonian approaches to quantum many-body systems}

To solve for the properties of nuclei, self-bound strongly interacting 
systems, with realistic Hamiltonians, one faces immense analytical and 
computational challenges.  Recently, {\it ab initio} methods have been 
developed that preserve all the underlying symmetries and 
converge to the exact result.  The basis function approaches that we adopt 
here, No Core Shell Model (NCSM)\cite{NCSM} and No Core Full Configuration 
(NCFC) methods \cite{NCFC}, are among the several methods shown to be 
successful.   The former adopts 
a finite basis-space renormalization method and applies it to realistic 
nucleon-nucleon (NN) and three-nucleon (NNN) interactions (derived from 
chiral effective field theory) to solve nuclei with Atomic Numbers 
$A =10-13$ \cite{chiral07}. Experimental binding energies, spectra, 
electromagnetic moments and transition rates are well-reproduced.
The latter adopts a realistic NN interaction that is sufficiently soft  that 
renormalization is not necessary and binding energies obtained from a sequence 
of finite matrix solutions may be extrapolated to the infinite matrix limit.   
Owing to uniform convergence and the variational principle, one is also able 
to assess the theoretical uncertainties in the extrapolated result.  One again 
obtains good agreement with experiment. 
The primary advantages of these methods are  the
flexibility for choosing the Hamiltonian, 
the method of renormalization/regularization and the basis space.  
These advantages direct us to adopt the basis function approach 
in light-front quantum field theory. 
\begin{figure}[h]
\includegraphics[width=8cm]{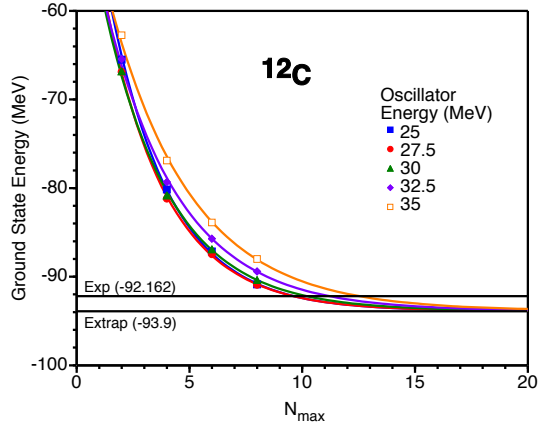}
\caption{ Calculated ground state energy of $^{12}$C for
$N_{max}=2{-}8$ (discrete points) at selected values of the oscillator energy, 
$\Omega$.  For each $\Omega$, the results are fit to an exponential plus
a constant, the asymptote, which is constrained to be the same for 
each curve\cite{NCFC}.  We display the experimental
ground state energy and the common asymptote. 
Figure from Ref.\cite{Vary:2009gt}.} 
\label{12C_JISP16_Jmax4_gs_vs_NMAX.eps}
\end{figure}
In a NCSM or NCFC application, one adopts a 3-D harmonic oscillator for all 
the particles in the nucleus (with harmonic oscillator energy $\Omega$), 
treats the neutrons and protons independently, and generates a many-fermion 
basis space that includes the lowest oscillator configurations as well as all 
those generated by allowing up to $N_{max}$ oscillator quanta of excitations.  
The single-particle states are formed by coupling the orbital angular momentum 
to the spin forming the total angular momentum $j$ and total angular momentum 
projection $M_j$.   The many-fermion basis consists of states where particles 
occupy the allowed orbits subject to the additional constraint that the total 
angular momentum projection $M_j$ is a pre-selected value. For the NCSM one 
also selects a renormalization scheme linked to the many-body basis space 
truncation while in 
the NCFC the renormalization is either absent or of a type that retains the 
infinite matrix problem. Non-perturbative renormalization has been 
developed to accompany these basis-space methods that preserve all the 
symmetries of the 
underlying Hamiltonian including highly precise treatments of the 
center-of-mass motion.  Several schemes have emerged with impressive successes
 and current research focuses on
detailed understanding of the scheme-dependence of convergence rates 
(different observables converge at different rates) \cite{Bogner07}.
In the NCFC case, one extrapolates to the continuum limit as illustrated in 
Fig.~\ref{12C_JISP16_Jmax4_gs_vs_NMAX.eps}, where
we show results for the ground state of $^{12}C$ as a function of $N_{max}$
obtained with a realistic NN interaction, JISP16 \cite{Shirokov07}.  
The smooth curves portray fits that achieve the desired independence of  
$N_{max}$ and $\Omega$ so as to yield the extrapolated ground state energy.   
Our assessed uncertainty in the extrapolant is about 2 MeV and there is rather 
good agreement with experiment within that uncertainty.  The largest cases 
presented in Fig.~\ref{12C_JISP16_Jmax4_gs_vs_NMAX.eps} correspond to 
$N_{max}=8$, where the matrix 
reaches a basis dimension near 600 million. Computation for the 
$N_{max}=10$ matrix case is in progress.
Large scale calculations like these
are performed on leadership-class parallel computers, 
at Argonne National Laboratory and at Oak Ridge National Laboratory, to solve 
for the low-lying eigenstates and eigenvectors as well as to carry out 
evaluation of a suite of experimental observables.  For example, one can now 
obtain the low-lying solutions for $A = 14$ systems with matrices of dimension 
one to three billion on 8000 to 50000 processors within a few hours of 
wallclock time.  Since the techniques are evolving rapidly \cite{Sternberg08} 
and the computers are growing dramatically, much larger matrices are within 
reach.

\section{Basis Light Front Quantized approach}

To take full advantage of the similarities with non-relativistic quantum 
many-body theory in what we will term a ``Basis Light Front Quantized 
(BLFQ)" approach, we adopt a light-front single-particle basis space 
consisting of the 2-D harmonic oscillator for the transverse modes (radial 
coordinate $\rho$ and polar angle $\phi$) and a discretized momentum space 
basis for the longitudinal modes.  Adoption of this basis is also consistent 
with recent developments in AdS/CFT correspondence with QCD 
\cite{Karch:2006pv,Erlich:2005qh,deTeramond:2008ht,Brodsky:2003px,Polchinski:2001tt}.  We define our light-front coordinates as
$x^{\pm}=x^0 \pm x^3$, $x^{\perp}=(x^1,x^2)$ and coordinate pair $(\rho,\phi)$
 are the usual cylindrical coordinates in $(x^1,x^2)$.  The variable $x^+$ 
is light-front time and $x^-$ is the longitudinal coordinate.  We adopt 
$x^+=0$, the ``null plane", for our quantization surface.

The 2-D oscillator states are characterized by their principal quantum number 
$n$, orbital quantum number $m$ and harmonic oscillator energy $ \Omega $.   
It is also convenient to interpret the 2-D oscillator as a function of the 
dimensionless radial  length scale variable  $\sqrt{M_0\Omega}\rho$ where 
$M_0$ has units of mass and $\rho$ is the conventional radial variable in 
units of length. 
The properly orthonormalized wavefunctions, 
$\Phi_{n,m}(\rho, \phi) = \langle \rho \phi | n m \rangle \ = 
f_{n,m}(\rho) \chi_{m}(\phi)$, are given in terms of the Generalized Laguerre 
Polynomials, $L_n^{|m|}(M_0\,\Omega\,\rho^2)$, by
\begin{eqnarray}
 && f_{n,m}(\rho) = \sqrt{2\,M_0\,\Omega} \sqrt{\frac{n!}{(n+|m|)!}}
   {\rm e}^{-M_0 \Omega \rho^2 / 2} \nonumber \\
  &&\hspace{1.5cm} \times\left(\sqrt{M_0\,\Omega}\rho\right)^{|m|} 
   L_n^{|m|}(M_0\Omega\rho^2) \nonumber \\
\label{Eq:wfn2dHOfx}
&&  \chi_{m}(\phi)=\frac{1}{\sqrt{2\pi}} {\rm e}^{i m\,\phi}
\label{Eq:wfn2dHOchix}
\end{eqnarray}
with eigenvalues $E_{n,m}=(2n+|m|+1)\Omega$.
In this 2-D oscillator basis 
the Fourier transformed wavefunctions have the 
same analytic structure in both coordinate and momentum space, a feature 
reminiscent of a plane-wave basis.

\begin{figure}[h]
\includegraphics[width=8cm]{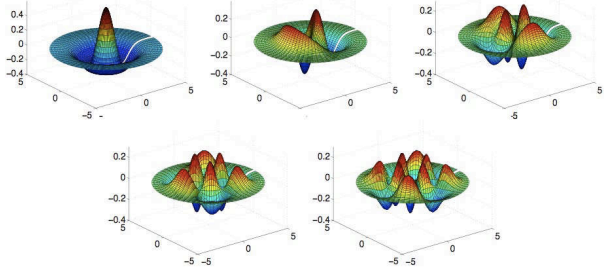}
\caption{Modes for $n=1$ of the 2-D harmonic oscillator 
selected for the transverse basis functions. The orbital quantum number $m$ 
progresses across the rows by  integer steps from 0 in the upper left to 4 in 
the lower right and counts the pairs of angular lobes.
Figure from Ref.\cite{Vary:2009gt}. } 
\label{fig3_paper.eps}
\end{figure}

\begin{figure}[h]
\includegraphics[width=8cm]{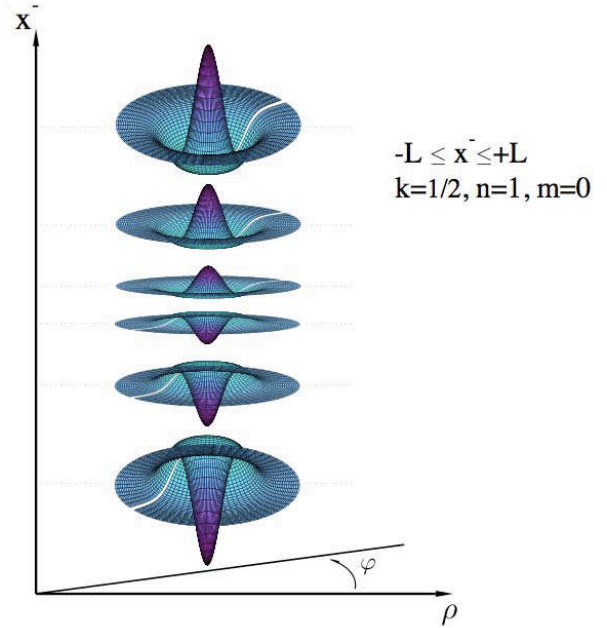}
\caption{Transverse sections of the real part of a 3-D basis 
function involving a 2-D harmonic oscillator and a longitudinal mode of 
Eq.(\ref {Eq:longitudinal1}) with antiperiodic boundary conditions (APBC). 
The quantum numbers for this basis function are given in the caption. The 
basis function is shown for the full range $-L \le x^- \le L$.
Figure from Ref.\cite{Vary:2009gt}.} 
\label{jun1.eps}
\end{figure}

The longitudinal modes, $\psi_{j}$, in our basis are  defined for
$-L \le x^- \le L$ with both periodic boundary conditions (PBC) and 
antiperiodic boundary conditions (APBC):
\begin{eqnarray}
  \psi_{k}(x^-) &=& \frac{1}{\sqrt{2L}} \, {\rm e}^{i\,\frac{\pi}{L}k\,x^-},
\label{Eq:longitudinal1}
\end{eqnarray}
where $k=1,2,3,...$  for PBC (neglecting the zero mode) and 
$k=\frac{1}{2},\frac{3}{2},\frac{5}{2},...$ in Eq.(\ref{Eq:longitudinal1}) for
APBC. The full 3-D single particle basis state is defined by the product form
\begin{eqnarray}
  \Psi_{k,n,m}(x^-,\rho,\phi) &=& \psi_{k}(x^-) \Phi_{n,m}(\rho, \phi).
\label{Eq:totalspwfn}
\end{eqnarray}
In Figs.~\ref{fig3_paper.eps} and \ref{jun1.eps} we illustrate some of
our basis functions. Fig.~\ref{fig3_paper.eps} presents modes for $n=1$ of 
the 2-D 
harmonic oscillator, and Fig.~\ref {jun1.eps} shows the 
transverse mode with $n=1, m=0$ joined together with the $k=\frac{1}{2}$ 
longitudinal APBC mode of Eq.(\ref{Eq:longitudinal1}) and display slices of the
 real part of this 3-D basis function at selected longitudinal coordinates, 
$x^-$. As one increases the orbital quantum number $m$,  
pairs of maxima and minima populate the angular dependence of the basis 
function.  Also, as one increases the principal quantum number $n$, 
additional radial nodes appear. 
As explained in \cite{Vary:2009gt} and references therein, our choice of 
basis functions is supported by the  the phenomenological success of the 
``soft-wall"  AdS/QCD model~\cite{Karch:2006pv,Erlich:2005qh} which uses  
a harmonic oscillator potential in the fifth dimension of Anti-de Sitter 
space to simulate color confinement.
The AdS/QCD model, together with light-front holography  
\cite{deTeramond:2008ht,Brodsky:2003px,Polchinski:2001tt},  provides  a 
semiclassical first approximation to strongly coupled QCD. The BLFQ approach 
in this paper provides a natural extension of the AdS/QCD light-front 
wavefunctions to multiquark and multi-gluonic Fock states, thus allowing 
for particle creation and absorption. By setting up and diagonalizing the 
light-front QCD Hamiltonian on this basis,
we incorporate higher order corrections corresponding to the full QCD theory,
and we hope to gain insights into the success of the AdS/QCD model.

\section{Cavity mode light-front field theory without interactions}

For a first application of the BLFQ approach, we consider a non-interacting 
QED system confined to a transverse harmonic trap or cavity.  The basis 
functions are matched to the trap so the length scale of the 2-D harmonic 
oscillator basis is fixed by the trap and finite modes in the 
longitudinal direction are taken with PBC from Eq.(\ref{Eq:longitudinal1}) (we omit the 
zero mode). Since we are 
ultimately interested in the self-bound states of the 
system, we anticipate adoption of the NCSM method \cite{NCSM}
for factorizing the 
eigensolutions into simple products of intrinsic and total momentum solutions 
in the transverse direction.  That is, with a suitable transverse
momentum constraint such as a large positive Lagrange multiplier times the 
2-D harmonic oscillator Hamiltonian acting on the total transverse 
coordinates, the low-lying physical solutions will all have the same 
expectation value of total transverse momentum squared.  Therefore, following 
Ref.\cite{BPP98} we introduce the total invariant mass-squared $M^2$ for 
these low-lying physical states in terms of a Hamiltonian $H$ times a 
dimensionless integer for the total light front momentum $K$
\begin{eqnarray}
M^2 + P_{\perp}P_{\perp} \rightarrow  M^2 + const = P^+P^- = KH
\label{Mass-squared}
\end{eqnarray}
where we absorb the constant into $M^2$.  
The non-interacting Hamiltonian $H_0$ for this system is defined by the sum 
of the occupied 
modes $i$ in each many-parton state with the scale set by the combined 
constant $\Lambda^2 = 2M_0\Omega$:
\begin{eqnarray}
&&  H_0 = 2M_0 P^-_c \nonumber \\
&&= \frac{2M_0\Omega}{K}\sum_i{\frac{2n_i+|m_i| +1 +
{{\bar m_i}^2}/(2M_0\Omega)
}{x_i}},\nonumber \\
\label{Hamiltonian}
\end{eqnarray}
where ${\bar m_i}$ is the mass of the parton $i$.

We adopt symmetry constraints and two cutoffs for our many-parton states. 
For symmetries, we fix the total charge $Z$, the total azimuthal quantum 
number $M_t$, and the total spin projection $S$ along the $x^-$ direction.  
For cutoffs, we select the total light-front momentum, $K$, and the maximum 
total quanta allowed in the transverse mode of each many-parton state, 
$N_{max}$. The chosen symmetries and cutoffs are expressed 
in terms of sums over the quantum numbers of the single-parton degrees of 
freedom contained in each many-parton state of the system in the following way:
\begin{eqnarray}
\sum_i{q_i} = Z
\\
\sum_i{m_i} = M_t
\\
\sum_i{s_i} = S
\\
\sum_i{x_i} = 1= \frac{1}{K}\sum_i{k_i} \label{x_def}
\\
\sum_i{2n_i+|m_i| +1} \le N_{max}
\end{eqnarray}
where, for example, $k_i$ is the integer that defines the PBC longitudinal 
modes of Eq.(\ref{Eq:longitudinal1}) for the $i^{th}$ parton. The range of the 
number of fermion-antifermion pairs and bosons is limited by the cutoffs in 
the modes ($K$ and $N_{max}$).  Since each parton carries at least one unit 
of longitudinal momentum,  the basis  is limited to $K$ partons.  Furthermore, 
since each parton carries at least one oscillator quanta for transverse 
motion, the basis is also limited to $N_{max}$ partons.  Thus the combined 
limit on the number of partons is  $\min({K},{N_{max}})$.  Only the case with 
simultaneous increase in both of these cutoffs keeps the problem physically 
interesting at higher excitations since this is the only case with unlimited 
number of partons as both cutoffs go to infinity. 
In principle, one 
may elect to further truncate the many-parton basis by limiting the number of 
fermion-antifermion pairs and/or the number of bosons but we have not elected 
to do so here.

In a fully interacting application, the actual choice of symmetry constraints 
will depend on those dictated by the Hamiltonian.  For example, with QCD we 
need to add color and flavor attributes to the single particle states and 
apply 
additional symmetries such as requiring all many-parton states to be global 
color singlets as discussed below.  Another example is the choice to conserve 
total $M_t + S$ rather than conserving each separately as chosen here.  It is 
straightforward, but sometimes computationally challenging, to modify the 
symmetries in a basis function approach such as we adopt here.  However, in 
order to approach the continuum limit (all cutoffs are removed) as closely as 
possible with limited computational resources, one works to implement as many 
of the known symmetries as possible.

\begin{figure}[h]
\includegraphics[width=8cm]{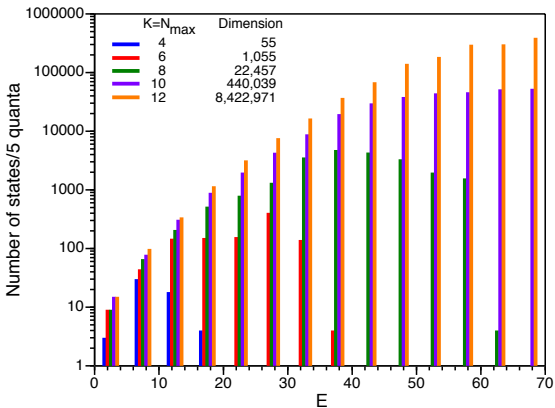}
\caption{State density as a function of dimensionless state 
energy E from BLFQ for non-interacting QED in a trap with no net charge and 
for $K=N_{max}$.  The dimensions of the resulting matrices are presented in 
the legend.  
The states are binned in groups of 5 units of energy (quanta) 
where each parton carries energy equal to its 2-D oscillator quanta 
($2n_i+|m_i|+1$) divided by its light-front momentum fraction ($x_i=k_i/K$). 
Figure from Ref.\cite{Vary:2009gt}.}
\label{Beq0_histogram_KNdep}
\end{figure}

In Fig.\ref{Beq0_histogram_KNdep} we illustrate how the BLFQ basis-space 
dimensions rapidly increase as a function of dimensionless state energy  
for non-interacting QED with massless partons.  The states are grouped to form 
a histogram according to their energy calculated from the chosen Hamiltonian 
in Eq.(\ref{Hamiltonian}) where we omit the constant preceding the summation 
for simplicity. We set $K = N_{max}$, 
and increase them simultaneously.
For simplicity we consider a case with no net charge $Z=0$, i.e. for zero 
lepton number.  Thus the cavity is populated by many-parton states consisting 
of fermion-antifermion pairs and photons.   The chosen symmetries are $M=0$ 
and $S=0$. As seen from Eqs.(\ref{Hamiltonian},\ref{x_def}),
fixing $K$ but increasing $N_{max}$ leads to a situation where partons carrying
large longitudinal momentum fraction become rare, and thus the low-lying modes 
get eventually maximally populated, leading to saturation.
 As shown in Ref.\cite{Vary:2009gt},
the energy at which this saturation occurs, increases with $N_{max}$.
In the case shown here,  there is no saturation in state density at low energy.
With increasing cutoff, there is a rapid growth in the number of basis states 
within each Fock space sector.  Overall, there is approximately a factor of 20 
increase in the total many-parton basis states with each increase of 2 units 
in the cutoff.
For the non-interacting  Hamiltonian the scale is available through an 
overall factor 
$\Lambda^2 = 2M_0\Omega$ as described above.  Without interactions and the 
associated renormalization program, one cannot relate the scales at one set 
of $(K,N_{max})$ values to another.  Ultimately, one expects saturation will 
arise with interaction/renormalization physics included as one increases the 
set of $(K,N_{max})$ values.

These state densities could serve as input to model the statistical mechanics 
of the system treated in the microcanonical ensemble.  Of course, interactions 
must be added to make the model realistic at low temperatures where 
correlations are important.  After turning on the interactions, the challenge 
will be to evaluate observables and demonstrate convergence with respect to 
the cutoffs ($N_{max}$ and $K$).  Independence of the basis scale, $\Omega$, 
must also be obtained as $N_{max}$ increases Ref.\cite{SCOON}.
These are the standard challenges of taking the 
continuum limit.

\begin{figure}[h]
\centering
\includegraphics[width=6cm]{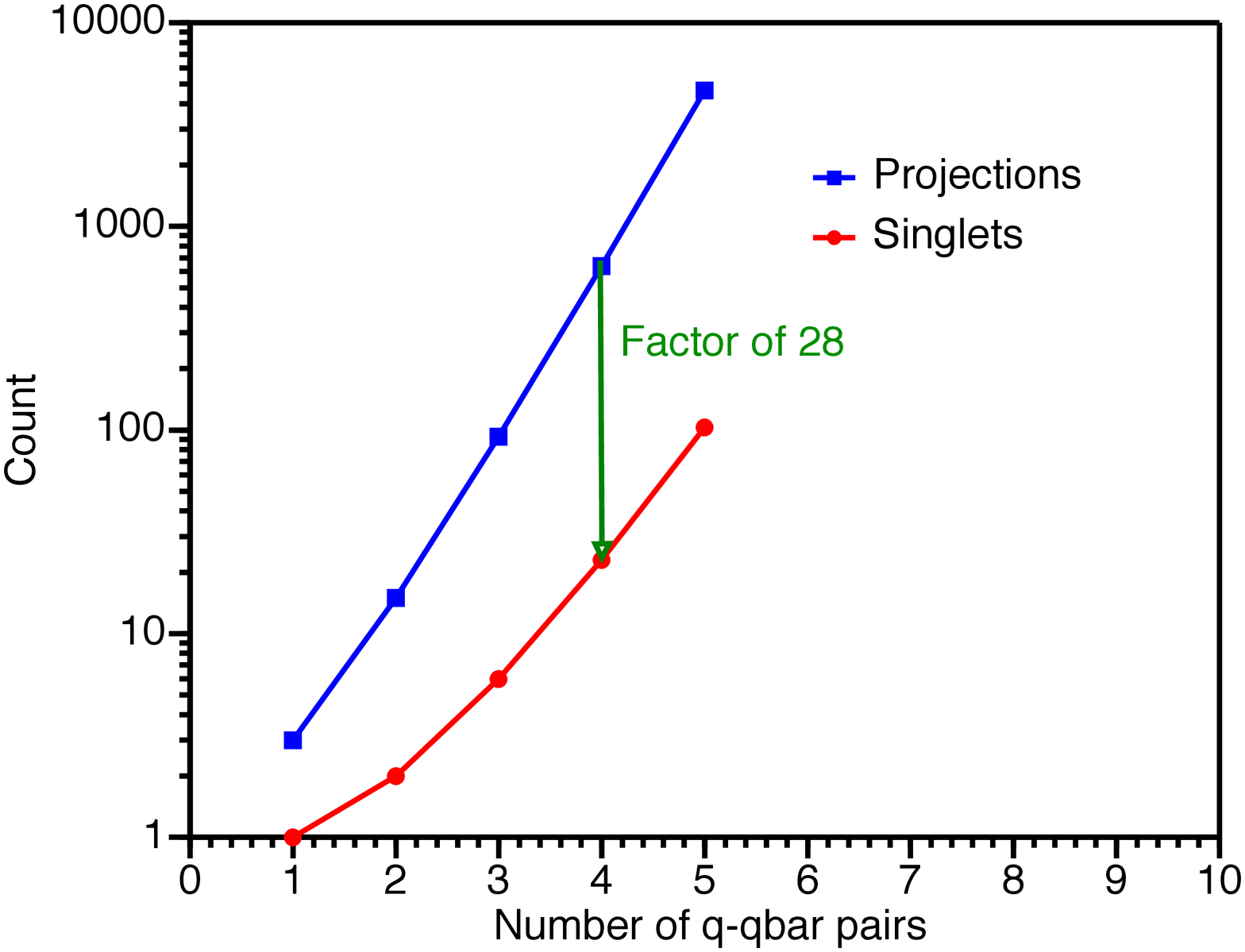}
\includegraphics[width=6cm]{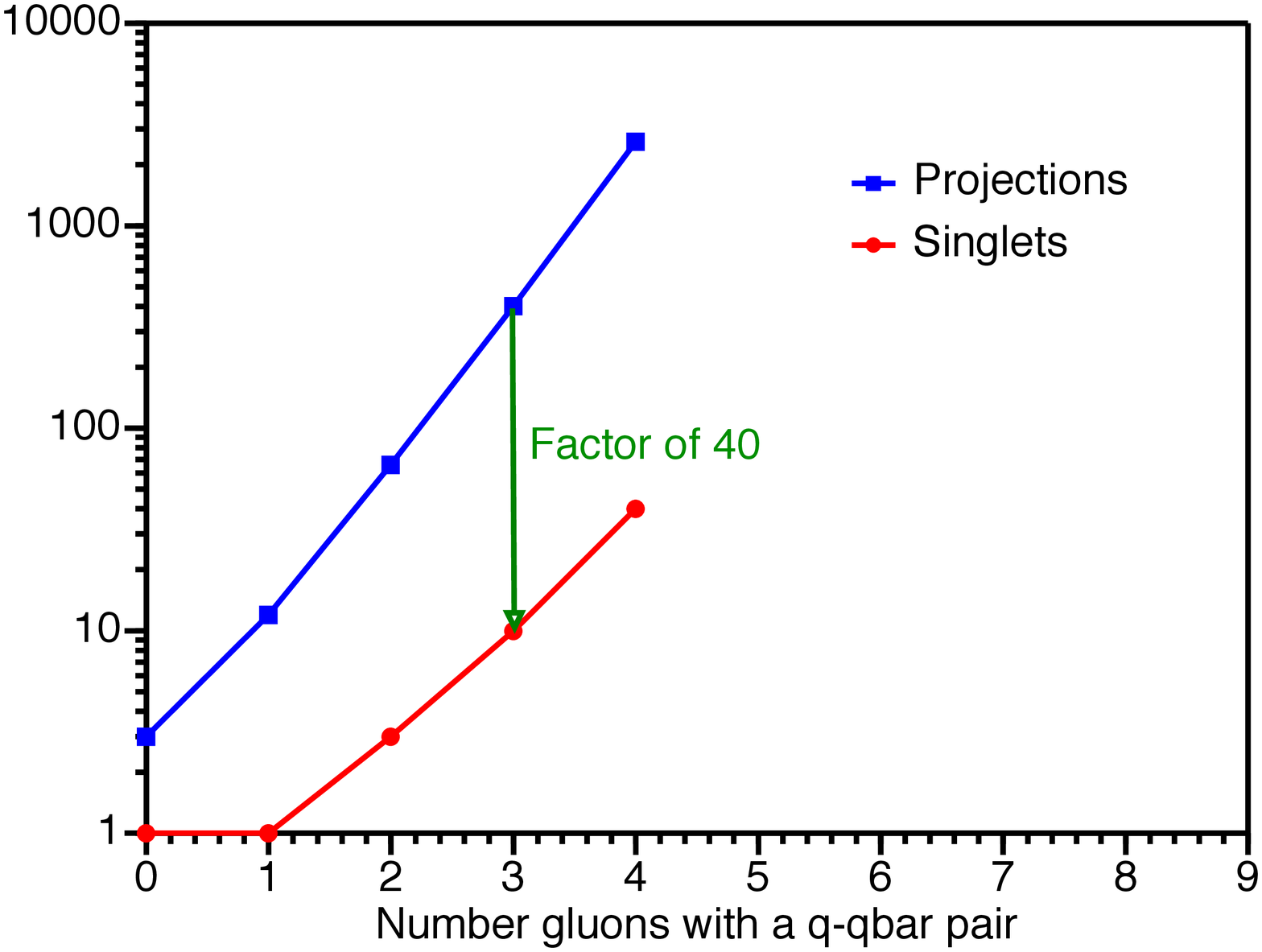}
\caption{Number of color space states that apply to each 
space-spin configuration of selected multi-parton states for two methods of 
enumerating the color basis states. The upper curves are counts of all color 
configurations with zero color projection.  The lower curves are counts of 
global color singlets. Figures from Ref.\cite{Vary:2009gt}.} 
\label{colorstates}
\end{figure}

The approach can be extended to QCD by implementing the SU(3) color degree of 
freedom for each parton - 3 colors for each fermion and 8 for each boson.  
For simplicity, we restrict the present discussion to the situation where 
identical fermions occupy distinct space-spin single-particle modes.  
We consider two versions of implementing the global color-singlet constraint 
for the restricted situation under discussion here.   In both cases we 
enumerate the color space states to integrate with each space-spin state of 
the corresponding partonic character. 
In the first case, we follow Ref. \cite{Lloyd} by enumerating parton states 
with all possible values of SU(3) color. Thus each space-spin fermion state 
goes over to three space-spin-color states.  Similarly, each space-spin boson 
state generates a multiplicity of eight states when SU(3) color is included. 
We then construct all many-parton states having zero color projection.  Within 
this basis one will have both global color singlet and color non-singlet 
states.  The global color-singlet states are then isolated by adding a 
Lagrange multiplier term in many-parton color space to the Hamiltonian so that 
the unphysical color non-singlet states are pushed higher in the spectrum away 
from the physical color single states.  To evaluate the increase in basis space
dimension arising from this treatment of color, we enumerate the resulting 
color-singlet projected color space states and display the results as the 
upper curves in Fig. \ref{colorstates}.

In the second case, we restrict the basis space to global color singlets and 
this results in the lower curves in Fig. \ref{colorstates}.  The second method 
produces a typical factor of 30-40 lower multiplicity at the upper ends of 
these curves at the cost of increased computation time for matrix elements of 
the interacting Hamiltonian.  That is, each interacting matrix element in the 
global color-singlet basis is a transformation of a submatrix in the zero 
color projection basis. Either implementation dramatically increases the state 
density over the case of QED, but the use of a global color-singlet constraint 
is clearly more effective in minimizing the explosion in basis space states.

\section{Elements of the interacting theory}

In the interacting theory  a primary concern will be 
to manage the divergent structure of the theory.  There are two possible 
locations for divergences in a Hamiltonian basis function approach: (1) the 
matrix elements themselves diverge, or (2) the eigenvalues diverge as one or 
more cutoffs are removed.
In our planned cavity field theory application with interactions, we will 
manage these divergences with the help of suitable counterterms and boundary 
conditions.  The infrared divergences in 
light-front momentum arising in both the fermion to fermion-boson vertex as 
well as in the instantaneous fermion-boson interaction are expected to be 
well-managed by previously defined counterterms \cite{JPV214} suitably 
transcribed for the transverse basis functions we have adopted.  We anticipate 
this prescription will work since the longitudinal modes we adopt are similar 
to those used in Ref. \cite{JPV214}. Finally, we expect the transverse 
ultraviolet divergences to be suitably-managed with our basis function 
selection.  As an alternative scheme for comparison  
we also plan to adopt a second approach that involves a recently 
proposed sector-dependent coupling constant renormalization scheme 
\cite{Karmanov}.  Another alternative which we may adopt uses the Pauli-Villars
regulator \cite{Brodsky:1998hs}.

Since we are introducing a basis-function approach for the transverse degrees 
of freedom, we need to investigate convergence rates with increased cutoff of 
the transverse modes $N_{max}$. 
Here, in a simple set of examples, we outline
 how we can search for additional sources of divergence with the help of a 
perturbation theory analysis. For a first investigation, we have examined the 
behavior of various sets of 
matrix elements for the fermion to  fermion-boson vertex.   For the purpose of 
this investigation, we adopt periodic (antiperiodic) boundary conditions for 
the longitudinal modes of the bosons (fermions) and we hold spin projections 
fixed for initial and final states.  We then adopt specific values for the 
longitudinal momentum fractions observing the conservation rule.   The trends 
we examine should not be sensitive to the specific values adopted.

Consider the second order energy shift, $\Delta E$, induced on a single parton 
in the transverse mode ($n,m$) by its coupling $V$ to partons in higher energy 
transverse modes ($n',m'$): 
\begin{eqnarray}
\Delta E_{n,m} \approx \int{\frac{|\langle n,m |V| n',m' \rangle|^2}
{E_{n,m} - E_{n',m'}} \rho (\bar n')d \bar n'},
\end{eqnarray}
where 
\begin{eqnarray}
\bar n' = 2n' + |m'|
\\
E_{n',m'} = (\bar n' + 1)\Omega
\\
\rho (\bar n') = \bar n' + 1
\end{eqnarray}
and the degeneracy is taken into account in $\rho$.

Thus, according to perturbation theory, we expect a UV divergence if the 
matrix element falls off too slowly with increasing $\bar n'$.  In particular,
 if the matrix element falls approximately as
 $(\bar n')^{-\frac{1}{2}}$, then we expect a logarithmic divergence since the 
integrand will have a net $(\bar n')^{-1}$ dependence.   If the falloff is 
even slower then we encounter a more serious divergence.
 Another possible source of a log divergence could arise within the selected 
sum over $m'$ in which case $\rho$, the level density factor in the integrand, 
is unity. Then, if the matrix elements for fixed $n,n'$ are approximately 
constant with increasing $m'$, we again find a log divergence in the sum over 
$m'$.

\begin{figure}[h]
\includegraphics[width=8cm]{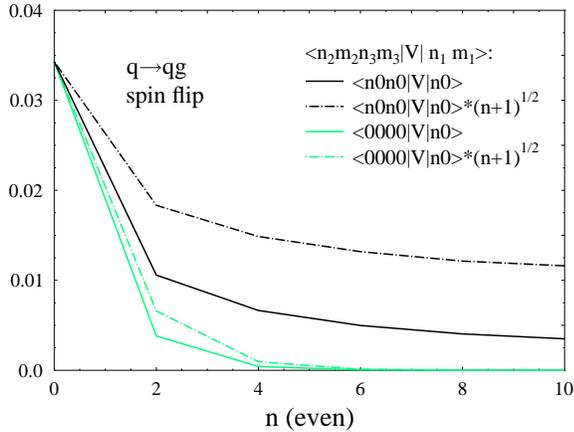}
\caption{Behavior of representative fermion to fermion-boson matrix elements 
in BLFQ. The quantum numbers specifying the parton transverse modes 
$(n_i,m_i)$ in the matrix elements are given in the legend. Only the 
transverse mode contributions to the matrix elements are shown.  Results are 
also shown with a multiplicative factor of $\sqrt{n+1}$ applied to help search 
for a logarithmic divergence by obtaining a resulting flat behavior, when it 
occurs. Overall matrix element normalization depends on the specific values 
of light-front momentum fractions carried by the interacting partons.
Figure from Ref.\cite{Vary:2009gt}. } 
\label{n_even}
\end{figure}

\begin{figure}[h]
\includegraphics[width=8cm]{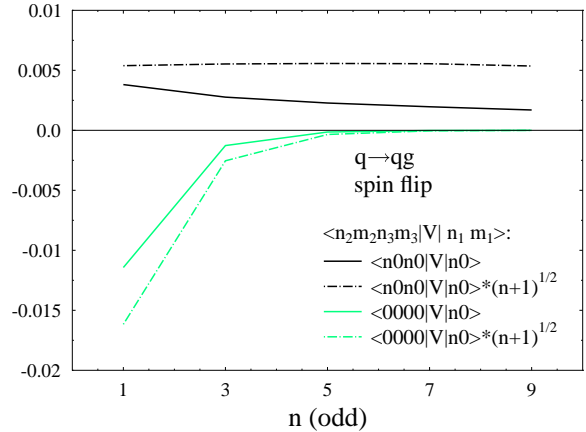}
\caption{ As in Fig.\ref{n_even}. Note that for one of the cases shown 
here, the resulting matrix elements vanish with 
increasing $n$ while the other case shows a constant trend but does not enter 
a second-order perturbative analysis. 
Figure from Ref.\cite{Vary:2009gt}.} 
\label{n_odd}
\end{figure}

We portray in Figs.\ref{n_even} and  \ref{n_odd} representative sequences of 
how these off-diagonal matrix elements behave as one increases the difference 
in the initial and final state principal quantum numbers.  We also portray two 
interesting cases where the fermion and fermion-boson principal quantum numbers
track each other,  cases that do not enter a perturbative analysis.  Note that 
we have limited the illustrations to the transverse components of our matrix 
elements. We also select  cases where the fermion spin is flipped, cases that 
are proportional to the fermion mass. We set the fermion mass to unity so the 
results are expressed in units of the fermion mass.
We further limit our presentation to the case where all partons remain in the 
orbital projection quantum number zero state and the 2-parton (fermion-boson) 
states have each parton in the same transverse state.  For the 
non-perturbative illustrative cases, we display the matrix element trends 
where all partons remain in the same transverse mode, $(n,0)$. 

When the single fermion state has an even value of the principal quantum 
number $n$ as shown in Fig.  \ref{n_even},  the matrix elements appear to be 
well-behaved either when the 2-parton configuration is the lowest accessible 
case ($\langle 0000 |$) or when the each of the two partons resides in the 
state with the same principal quantum number as the single fermion state.  We 
demonstrate anticipated good convergence with increasing $n$ by showing that 
the matrix elements, when multiplied by $\sqrt{n+1}$, still fall with 
increasing $n$.
For the case when the single fermion state has an odd value of the principal 
quantum number $n$ as shown in Fig.  \ref{n_odd}, the situation is somewhat 
different.  For the matrix element set entering a perturbative analysis, the 
matrix elements fall to zero with increasing $n$ sufficiently fast that 
multiplying by $\sqrt{n+1}$ does not significantly distort the trend to zero.  
However, the large $n$ behavior of the fermion-boson matrix element, with all 
partons at the same $n$, is seen to go approximately as $\sqrt{n+1}$.  This is 
best seen in Fig.  \ref{n_odd} where the matrix elements are multiplied by 
$\sqrt{n+1}$ and the result appears to be a nonzero constant at large $n$.  
Since this  trend does not appear in a second order perturbation theory 
analysis, we must await the full Hamitonian diagonalization to  better 
understand its role in the convergence with increasing $N_{max}$.
To date, we have found no matrix element trends with increasing transverse 
energy that would imply new divergences.  All such matrix elements sets we 
examined, within a perturbative analysis, fall faster than 
$1/\sqrt{n'}$.  
Also, we found no sets that 
remained constant with increasing $m'$ (and thus $m$) holding  $n,n'$ fixed. 
As a result of this initial analysis, we anticipate that straightforward 
adoption of counterterm methods previously introduced  \cite{JPV214} will be 
sufficient for managing the identified divergences in BLFQ.

\begin{figure}[h]
\includegraphics[width=9cm]{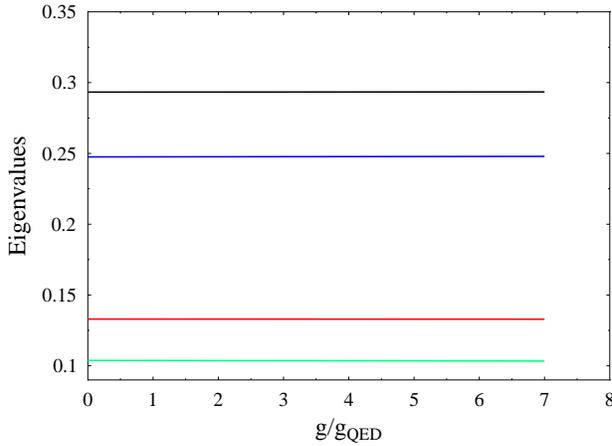}
\caption{ Eigenvalues of a cavity mode QED Hamiltonian, which includes lepton 
and lepton-photon Fock spaces, as a function of the strength of the coupling
($g_{\rm QED}=\sqrt{4\pi\alpha}$). The HO parameters are $\Omega=0.1$ MeV 
and $M_0=0.511$ MeV,  and the other
basis parameters are $K=3$, $N_{max}=2$ and $M_j=1/2$.} 
\label{eigenvalues_4x4.eps}
\end{figure}

As another example of the interacting cavity mode QED, we present  
in Fig.\ref{eigenvalues_4x4.eps} the  eigenvalue spectra for a Hamiltonian matrix
that includes lepton and lepton-photon Fock spaces. The leptons are now massive
with $\bar m_l=0.511$ MeV, and the HO parameters are $\Omega=0.1$ MeV and 
$M_0=\bar m_l$. The other basis parameters are chosen as $K=3$, $N_{max}=2$ 
and $M_j=1/2$. 
The eigenvalues are shown as a function of the strength of the coupling
with respect to the coupling at QED, with $g_{\rm QED}=\sqrt{4\pi\alpha}$.
Since the instantaneous fermion interaction term
only contributes when all the leptons and photons of the process have their
spins aligned, this particular case only involves the photon emission vertex
interaction (see e.g. Ref.\cite{Anommm}).
Application of the sector-dependent renormalization \cite{Karmanov}
will shift these eigenvalues such that the lowest one will obtain the value
$E_0=\frac{\bar m^2+M_0\Omega}{K}$, where $\bar m$ is the mass of the lowest
state fermion.
As we increase the cutoffs $K,N_{max}$, the evaluation of the electron
anomalous magnetic moment will serve as an important precision
test for our method
\cite{Anommm}.

\section{Conclusions}

Following successful methods of {\it ab initio} nuclear many-body theory, we 
have introduced a basis light-front quantization (BLFQ) approach to Hamiltonian
quantum field theory. With a cavity mode treatment of massless non-interacting 
QED we showed how the number of many-parton basis states exhibit  a dramatic 
rise as the cutoffs are elevated. In order to extend our method to QCD, we 
presented two methods for treating the color degree of freedom and 
demonstrated sample measures of the efficiency gains for global color singlets
over basis states with color-singlet projection alone.
For the  interacting  cavity mode QED we outlined our approach to managing 
the expected divergences in a manner that will preserve all the symmetries of 
the theory. An initial inspection of the 
interaction vertices of QED in the BLFQ approach shows smooth behaviors that, 
following a second-order perturbative analysis, are not expected to lead to 
divergences.  It appears that the cavity mode treatment, with the type of 
basis spaces we have selected, will encounter the divergences in a more subtle 
fashion as cutoffs are elevated. The evaluation of the electron
anomalous magnetic moment will be an important test to our method as we expand 
our basis space.

\end{document}